# Direct Conversion Pulsed UWB Transceiver Architecture

Raúl Blázquez, Fred Lee, David Wentzloff, Brian Ginsburg,
Johnna Powell, Anantha Chandrakasan
Massachusetts Institute of Technology
38-107, 50 Vassar St., Cambridge, MA 02139, USA
rbf@mit.edu

**Abstract**

*Ultra-wideband (UWB) communication is an emerging wireless technology that promises high data rates over short distances and precise locationing. The large available bandwidth and the constraint of a maximum power spectral density drives a unique set of system challenges. This paper addresses these challenges using two UWB transceivers and a discrete prototype platform.*

## 1. System considerations

The use of ultra-wideband signals for communication purposes is approved by the FCC from 3.1 to 10.6 GHz with a maximum effective isotropic radiated power spectral density of -41.3 dBm/MHz. An effective high data rate transceiver in this band should provide robust communication under severe multipath conditions (rms delay spread of the channel on the order of 20 ns) and narrowband interferers. In addition, a fast signal acquisition algorithm must be implemented to reduce the duration of the preamble to a value comparable with current wireless systems (~20 µs).

The RF Front-end must meet the specifications on noise figure and linearity over a bandwidth larger than 500 MHz. The impulse responses of both the antenna and the RF front-end add to that of the channel. Since the receiver will only be able to deal with a maximum channel impulse response set by design, the RF front-end must be designed to meet this constraint.

The ADC requires sampling rates larger than 500 MSps. The specification of the data converter resolution determines not only its power dissipation but also that of the digital back end. A 1-bit analog-to-digital converter (ADC) in a noise limited regime, and a 4-bit ADC in a narrowband interferer regime are sufficient [1].

The back end requires parallelization to reduce the packet synchronization time and to process the large data rate provided by the ADC. The inter-symbol interference (ISI) due to multipath can be addressed with a Viterbi demodulator. The energy spread caused by the multipath can be compensated using a RAKE receiver. These elements require an estimation of the impulse response that may be obtained during the packet synchronization. The frequency of an interferer may also be estimated in the digital back end. The large complexity required in the synchronization and demodulation of the UWB signal results in more than half of the system power being dissipated in the digital back end and the ADC.

## 2. First generation transceiver

Fig. 1 shows a system-on-a-chip transceiver for baseband pulsed UWB signals [2] that was implemented using a 0.18 µm CMOS technology at 1.8 V in a 4.3 mm × 2.9 mm die. A wireless link of 193 kbps was demonstrated with this transceiver. The RF front does not require a down converter. The analog-to-digital converter is a 4-way time-interleaved flash ADC that performs an initial 4-way parallelization of the signal. The timing synchronization is fully performed in the digital back end. Through further parallelization, packet synchronization is obtained in less than 70 µs.

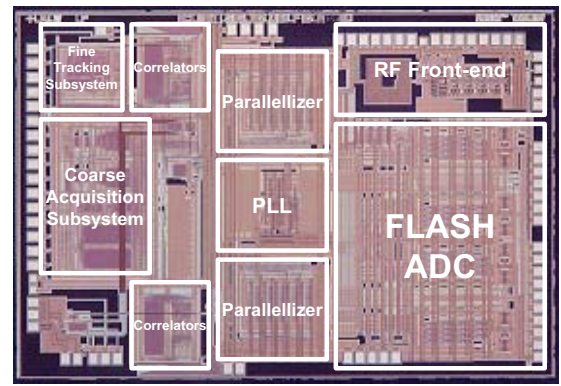

**Figure 1. Single chip UWB transceiver.**

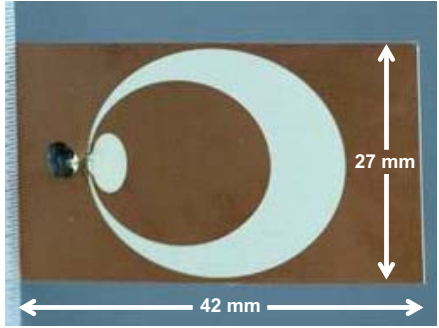

**Figure 2. Planar antenna for UWB.**

## 3. Architecture of a 3.1-10.6 GHz transceiver

We are currently developing a second generation UWB system to operate in the 3.1-10.6 GHz band. The signal is comprised of a sequence of 500 MHz bandwidth pulses that are upconverted to one of 14 channels (sub-bands) in the 3.1-10.6 GHz band. The system is designed to transmit 100 Mbps. An electrically small planar antenna [3] of only 4.2 cm × 2.7 cm has been developed (Fig. 2). The block diagram of the receiver is shown in Fig. 3. The RF front end uses a direct conversion architecture. Both in-phase and quadrature channels are then digitized using two 5-bit successive approximation register ADCs that feed the digital back end.

In order to cope with the multipath, the channel impulse response is estimated with a precision of up to four bits during the packet preamble. This information is used in a RAKE receiver and in a Viterbi demodulator, both of them programmable. The digital back end detects the presence of an interferer and estimates its frequency that may be used in the front end notch filter. This receiver allows us to trade off power dissipation with signal processing complexity, quality of service and data rate, adapting to channel conditions.

A discrete prototype with the same specifications has been designed and implemented, allowing not only the testing of the different parts of the transceiver for specifications, but also a complete testing of the algorithms implemented in the digital back end under realistic conditions. This platform is also flexible enough to generate all kinds of signals within a bandwidth of 500 MHz, allowing the comparison between different modulation schemes. Fig. 4 shows an example of a modulated pulse used in this discrete prototype.

## 4. Conclusions

The key system challenges for UWB communication systems are outlined in this paper. The design trade-offs are illustrated using two different UWB transceivers and a discrete platform.

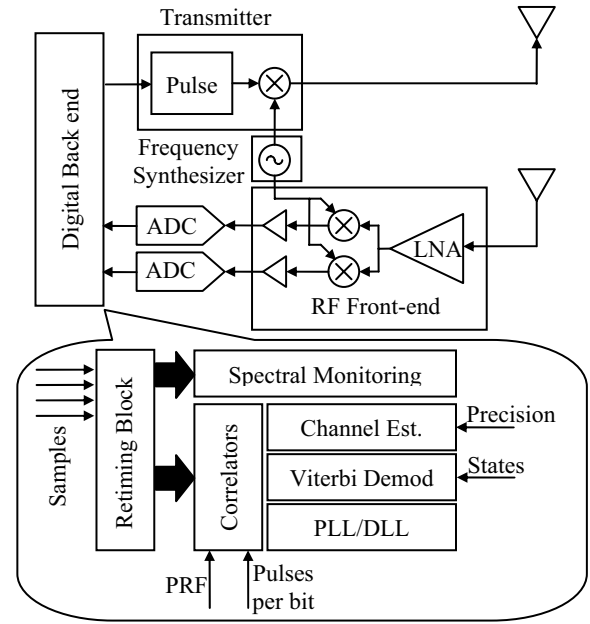

**Figure 3. UWB system block diagram.**

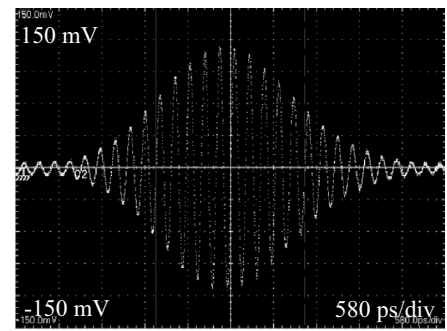

**Figure 4. 500 MHz pulse with carrier 5 GHz.**

## Acknowledgements

This research is sponsored by Hewlett-Packard under the HP/MIT Alliance and the NSF under contract ANI-0335256.